\documentclass[12pt]{amsart}

\usepackage[centertags]{amsmath}
\usepackage{amsfonts}
\usepackage{amssymb}
\usepackage{amsthm}
\usepackage{newlfont}
\usepackage{epsfig}
\usepackage{amscd}

\newcommand{\bD}{{\boldsymbol D}}
\newcommand{\bE}{{\boldsymbol E}}
\newcommand{\bA}{{\boldsymbol A}}

\newcommand{\PP}{{\mathbb P}}
\newcommand{\CC}{{\mathbb C}}
\newcommand{\ZZ}{{\mathbb Z}}
\newcommand{\KK}{{\mathbb K}}
\newcommand{\FF}{{\mathbb F}}
\newcommand{\LL}{{\mathbb L}}
\newcommand{\C}{{\mathcal C}}
\newcommand{\calC}{{\mathcal C}}

\newtheorem*{Lem}{Lemma}
\newtheorem*{Prop}{Proposition}

\theoremstyle{definition}
\newtheorem*{Def}{Definition}
\newtheorem{Ex}{Example}

\theoremstyle{remark}
\newtheorem*{Rem}{Remark}
\begin{document}

\title[Integrable cellular automata]{Integrable 1D Toda cellular automata}
\author[M. Bia{\l}ecki]{Mariusz Bia{\l}ecki}
\address{M. Bia{\l}ecki \newline 
Institute of Geophysics, Polish Academy of Sciences\\ \newline 
ul. Ksi\c{e}cia Janusza 64\\ 01-452 Warszawa\\ Poland\\}
\email{bialecki@igf.edu.pl} 

\begin{abstract}
First, we recall the algebro-geometric method of construction of 
finite field valued solutions  
$$\ZZ^3 \ni(n_1,n_2,n_3) \longmapsto \tau(n_1,n_2,n_3) \in \FF_q,$$
of the {\bf discrete KP} equation ($T_i$ --- shift operator in $n_i$ variable)
\begin{equation*}  \label{eq:alg-tauKP} 
(T_1\tau) \cdot (T_2T_3\tau) - (T_2\tau) \cdot (T_3T_1 \tau) + 
(T_3\tau) \cdot (T_1T_2\tau) = 0, 
\end{equation*} 
and next we perform a reduction of dKP to the {\bf discrete 1D Toda} equation 
\begin{equation*} (T_1^{-1}T_3\tau) \cdot (T_1T_3^{-1}\tau) - (T_1^{-1}\tau) \cdot ( T_1 \tau) + 
\tau \cdot \tau  = 0. 
\end{equation*} 
This gives a method of construction of solutions of the discrete 1D Toda 
equation taking values in the finite field $\FF_q$.
\end{abstract}

\maketitle

\section{Introduction}

Discrete integrable systems focus much attention for at least two reasons.
They inherit developments of the theory of continuous integrable models on the one hand,
and give rise to new ideas on the other. 
Recently  this "twofold" nature was exploited for the construction 
of integrable cellular automata. 

The approach is based on transferring the standard algebro-geometric method of construction of solutions 
\cite{Krich-discr,BBEIM} from the complex field $\CC$ to a finite field case. 
In spite of apparent differences between complex and finite fields (see \cite{LidlNied}) 
the theory of algebraic functions over finite field \cite{Goldschmidt, Sticht}
provides a powerful framework and makes this transfer possible. As a consequence of 
{\emph{algebraic}} approach it appears that relevant theorems (as the Riemann-Roch theorem)
and formulas are "quite the same" as in the complex field case but have a different meaning.
For instance, the concept of a continuum limits (in expansions etc.) is not clear 
as far as finite fields are regarded. 

For a given difference equation, its finite field valued solution defines a cellular automaton. 
So far there have been constructed multisoliton solutions for the discrete 2D Toda lattice 
(the Hirota's bilinear difference equation) \cite{DBK}, 
the discrete Kadomtsev-Petviashvili (dKP) and discrete Korteweg-de Vries (dKdV) equations \cite{BD-KP}
and also algebro-geometric solutions for dKP \cite{BD-hyp} and dKdV \cite{Bia-hyp}. 
In this article we extend previous results for the discrete 1D Toda equation 
(and ipso facto for a certain discrete version of the sine-Gordon equation, 
see remark at the end of section \ref{rem:dSG}).    

The paper consists of three main parts. 
In Section \ref{sec:dKP} we recall the finite field version of the algebro-geometric construction of
solutions of the discrete KP equation. 
In Section \ref{sec:1dT} the algebro-geometric reduction to the discrete 1 dimensional Toda equation 
is performed and illustrated by examples in Section \ref{sec:Ex}.

\section{An algebro-geometric approach to dKP equation} \label{sec:dKP}

\subsection{General construction for the dKP equation} \label{sec:gen-constr}
For the general construction \cite{BD-KP} we need an algebraic projective curve 
$\calC/\KK$ (or simply $\calC$),  
absolutely irreducible, nonsingular, of genus $g$, defined over the finite  
field $\KK=\FF_q$ with $q$ elements.
$\calC(\KK)$ denotes the set of $\KK$-rational points of the curve. 
$\overline{\KK}$ denotes the algebraic closure of  
$\KK$, i.e., $\overline{\KK} = \bigcup_{\ell=1}^\infty \FF_{q^\ell}$, and  
$\calC(\overline{\KK})$ denotes the corresponding infinite set of 
$\overline{\KK}$-rational points of the curve. 
Denote by $\mathrm{Div}(\calC)$ the abelian group of the divisors on the curve $\calC$.
The action of the Galois group $G(\overline{\KK}/\KK)$ (of automorphisms of  
$\overline{\KK}$ which are identity on $\KK$) extends 
naturally to action on $\calC(\overline{\KK})$ and $\mathrm{Div}(\calC)$. 
A field of $\KK$-rational functions on the curve $\calC$ is denoted by $\KK(\calC)$. 
A vector space $L(D)$ is defined as $\{ f \in \overline\KK(\calC) \ | \ (f) > -D \}$,
where $D \in \mathrm{Div}(\calC)$ and $(f)=\sum_{P\in \calC} ord_P (f)\cdot P$ \ 
is the divisor of the function $f\in\KK(\calC)$.

On the curve $\calC$ we choose:
\begin{enumerate} 
\item four points $A_0, A_i\in\calC(\KK)$, \  $i=1,2,3$,  
\item effective $\KK$-rational  divisor of order $g$,
i.e., $g$ points $B_\gamma\in\calC(\overline{\KK})$, $\gamma=1,\dots,g$, 
which satisfies the following $\KK$-rationality condition 
\[ 
\forall \sigma\in  G(\overline{\KK}/\KK), \quad  
\sigma(B_\gamma) = B_{\gamma^\prime},
\]
\item{effective $\KK$-rational  divisor of order $N$,
i.e. $N$ points $C_\alpha\in\cal C({\overline{\KK}})$, $\alpha=1,\dots,N$, 
 that satisfies the $\KK$-rationality conditions   
\[ 
\forall \sigma\in  G(\overline{\KK}/\KK), \quad  
\sigma(C_\alpha) = C_{\alpha^\prime}, 
\] }
\item{$N$ pairs of points $D_\beta, E_\beta\in\cal C({\overline{\KK}})$,  
$\beta=1,\dots,N$, that satisfy $\KK$-rationality conditions  
\begin{equation*} \label{eq:K-rat-cond} 
\forall \sigma\in  G(\overline{\KK}/\KK): \quad 
\sigma(\{D_\beta,E_\beta\})=\{D_{\beta^\prime},E_{\beta^\prime}\}. 
\end{equation*} 
}
\end{enumerate}
We assume that all the points used 
are distinct and in general position. In particular, 
the divisors $\sum_{\gamma=1}^g B_\gamma$, $\sum_{\alpha=1}^N C_\alpha$,
and  $D(n_1,n_2,n_3)$, defined below, are non-special. 

\begin{Def}  \label{def:psi}
Fix $\KK$-rational local parameter $t_0$ at $A_0$. 
For any integers $n_1,n_2,n_3\in \ZZ$ let divisor $D(n_1,n_2,n_3)$ be of the form
$$D(n_1,n_2,n_3)= \sum_{i=1}^3 n_i(A_0-A_i)+ \sum_{\gamma=1}^g B_{\gamma} +
 \sum_{\alpha=1}^N C_{\alpha}.$$
The function  
$\psi(n_1,n_2,n_3)$ (called a wave function) is a 
rational function on the curve $\calC$ with the following properties
\begin{enumerate}
\item the divisor of the function satisfies $(\psi)> -D$, i.e. $\psi \in L(D)$,  
\item{ the function $\psi$ satisfies  $N$  constraints
\begin{equation*} \label{eq:constraints} 
\psi(n_1,n_2,n_3)(D_\beta)=\psi(n_1,n_2,n_3)(E_\beta), \quad 
\beta=1,\dots,N. 
\end{equation*} }  
\item the first nontrivial coefficient of its expansion in $t_0$ at $A_0$ is 
normalised to one. \label{point-norm}
\end{enumerate}
\end{Def} 
Existence and uniqueness of the function $\psi(n_1,n_2,n_3)$ is due 
to application of the Riemann--Roch theorem with general position
assumption and due to normalisation. 
Moreover, the function $\psi(n_1,n_2,n_3)$ is $\KK$-rational, which follows  
from $\KK$-rationality conditions for sets of points in their definition.

The next step of the construction is to obtain linear equations
for wave functions. The full form of such equation is in case when 
the pole of $\psi(n_1,n_2,n_3)$ at $A_0$ is of the order exactly $(n_1+n_2+n_3)$ 
and respective zeros at $A_i$ are of the order $n_i$, for $i=1,2,3$. 
We will call this case generic.
Having fixed $\KK$-rational local parameters $t_i$ at $A_i$, $i=1,2,3$, 
denote by $\zeta^{(i)}_k(n_1,n_2,n_3)$, $i=1,2,3$, 
the $\KK$-rational
coefficients of expansion of $\psi(n_1,n_2,n_3)$ at $A_i$, respectively, 
i.e.,
\begin{align*} 
\psi (n_1,n_2,n_3) &= t_i^{n_i} \sum_{k=0}^{\infty} 
\zeta^{(i)}_k(n_1,n_2,n_3)  t_i^k , \quad i=1,2,3.  
\end{align*} 
Denote by $T_i$ the operator of translation 
in the variable $n_i$, $i=1,2,3$, 
for example, $T_2 \psi(n_1,n_2,n_3) = \psi(n_1,n_2+1,n_3) $. 
The full linear equation is of the form
 \begin{equation}  
T_i \psi - T_j \psi +  
\frac{T_j\zeta^{(i)}_ 0}{\zeta^{(i)}_0} \psi = 0 , \quad i\ne j , 
\quad i,j=1,2,3. 
\label{eq:psi}  
\end{equation} 
It follows from observation that $T_i \psi - T_j \psi \in L(D)$, hence it must be
proportional to wave function $\psi$. Coefficients of proportionality can be 
obtained from comparison (the lowest degree terms) of expansions of left and 
right sides of \eqref{eq:psi} at the point $A_i$.

\begin{Rem}
When the genericity assumption fails then the linear problem~\eqref{eq:psi} 
degenerates to the form $T_i\psi=\psi$ or even to $0=0$. 
\end{Rem}

Notice that equation \eqref{eq:psi} gives
\begin{equation} \label{eq:zeta-zeta}
\frac{T_j\zeta^{(i)}_ 0}{\zeta^{(i)}_0} =
 -  \frac{T_i\zeta^{(j)}_ 0}{\zeta^{(j)}_0} , \quad i\ne j,\quad i,j=1,2,3.  
\end{equation}
Define 
\begin{equation*} \label{eq:rho-def}
\rho_i= (-1)^{\sum_{j<i} n_j} \zeta^{(i)}_0, \quad i=1,2,3,
\end{equation*} 
then equation
\eqref{eq:zeta-zeta} implies existence of a $\KK$-valued potential 
(the $\tau$-function) defined (up to a multiplicative
constant) by formulas 
\begin{equation} 
 \frac{T_i\tau}{\tau} = \rho_i,\quad  i=1,2,3.  
\end{equation} 
Finally, by "cyclic" use of equations \eqref{eq:psi} and \eqref{eq:zeta-zeta} 
one can get the condition 
\begin{equation}  \label{eq:KP-rho}
\frac{T_2\rho_1}{\rho_1} - \frac{T_3\rho_1}{\rho_1} + 
\frac{T_3\rho_2}{\rho_2} = 0, 
\end{equation}  
which, written in terms of the $\tau$-function, gives the discrete KP
equation~\cite{Hirota} called also the Hirota equation
\begin{equation}  \label{eq:tauKP} 
(T_1\tau) \;(T_2T_3\tau) - (T_2\tau) \;(T_3T_1 \tau) + 
(T_3\tau) \;(T_1T_2\tau) = 0. 
\end{equation} 
\begin{Rem} 
Equation \eqref{eq:KP-rho} can be obtained also from expansion of equation 
\eqref{eq:psi} at $A_k$, where $k=1,2,3$, $k\ne i,j$.
\end{Rem}

Absence of a term in the linear problem \eqref{eq:psi}, due to the
Remark above, reflects in absence of
the corresponding term in equation \eqref{eq:tauKP}. This implies that in the
non-generic case, when we have not defined the $\tau$-function yet, we are 
forced to put it to zero. 

\subsection{Explicit formulas for projective line}

Denote by $t$ standard parameter on projective line $\PP(\KK)$ and choose  $A_0=\infty$.
Then we have the explicit formulas: vacuum wave functions
\begin{eqnarray} 
\psi^0 &= & 
{(t-A_1)^{n_1}} 
{(t-A_2)^{n_2}} {(t-A_3)^{n_3}} \nonumber 
\end{eqnarray}     
and vacuum $\tau$-functions 
\[ 
\tau^0 = {(A_1-A_2)^{n_1 n_2}} {(A_1-A_3)^{n_1 n_3}} {(A_2-A_3)^{n_2 n_3}}.    
\] 
Non vacuum  $\tau$-function can be constructed by (see \cite{DBK})
\begin{equation} \label{eq:tau-N-soliton} 
\tau = \tau^0 \det \phi^0_{\bA}(\bD,\bE), 
\end{equation}      
where we denote by $\phi^0_{\bA}(\bD,\bE)$ the 
$N\times N$ matrix whose element in row $\beta$ and  column $\alpha$ is 
\[ 
\left[\phi^0_{\bA}(\bD,\bE)\right]_{\alpha \beta}= 
\phi^0_{\alpha}(D_\beta)-\phi^0_{\alpha}(E_\beta) , \quad 
\alpha,\beta=1,\dots,N. 
\] 
and auxiliary vacuum wave functions $\phi^0_{\alpha}$, $\alpha=1,\dots,N$ 
have the form
\begin{equation*} \label{eq:phi-KP-vac} 
\phi_{\alpha}^0= \frac{1}{t-C_\alpha} \cdot 
\frac{(t-A_1)^{n_1} (t-A_2)^{n_2} (t-A_3)^{n_3}} 
{(C_\alpha-A_1)^{n_1} (C_\alpha-A_2)^{n_2}(C_\alpha-A_3)^{n_3}}. 
\end{equation*} 

\bigskip

\begin{Rem} \label{rem:periods}
Let $\LL=\FF_{q^l}$ be a field of rationality of all the points used in the construction. 
Then $\tau$-function is periodic in $n_1$,  $n_2$ and  $n_3$ with  periods being 
divisors of $q^l-1$, which is the order of the 
 multiplicative group $\LL_*$.
\end{Rem}

Algebro-geometric solutions can be obtained using the Jacobian of the curve $\C$.
Construction of KP cellular automata from hyperelliptic curves was presented in
\cite{BD-hyp} and of KdV cellular automata in \cite{Bia-hyp}.

\section{Reduction to the discrete 1D Toda equation} \label{sec:1dT}

\subsection{Reduction to the discrete 1D Toda equation}

The discrete 1D Toda equation 
\begin{equation} \label{eq:discrete 1D Toda}
(T_1T_3^{-1}\tau) \;(T_1^{-1}T_3\tau) - (T_1\tau) \;( T_1^{-1} \tau) + 
\tau \;\tau  = 0. 
\end{equation} 
can be obtained from the  dKP equation 
\begin{equation*} \label{eq:dKP} 
(T_1\tau) \;(T_2T_3\tau) - (T_2\tau) \;(T_3T_1 \tau) + 
(T_3\tau) \;(T_1T_2\tau) = 0. 
\end{equation*} 
by imposing the constraint
\begin{equation} \label{eq:KPtoSGconstr}
 T_1T_2T_3^{-1}\tau = \gamma \tau,
\end{equation}
where $\gamma$ is a non-zero constant.
The reduction 
\eqref{eq:KPtoSGconstr} can be expressed at the level of the 
wave function $\psi$ as follows.  

\begin{Lem} \label{lem:KP-to-SG}
Assume that on the algebraic curve $\calC$ there exists a rational
function $h$ with the following properties
\begin{enumerate}
\item the divisor of the function is $(h)=-A_0+A_1+A_2-A_3$,
\item it satisfies $N$ constraints: $h(D_\beta)=h(E_\beta)$,
$\beta=1,\dots,N$,
\item the first nontrivial coefficient of 
its expansion in the parameter $t_0$ at $A_0$ is normalised to one. \label{point-norm-h}
\end{enumerate}
Then the wave function $\psi$ satisfies the following condition
\begin{equation*} \label{eq:KPtoSG-constr-psi}
T_1T_2T_3^{-1}\psi = h \psi.
\end{equation*}
\end{Lem}

\begin{proof}
Orders of poles and zeros of functions 
from both sides of equation \eqref{eq:KPtoSG-constr-psi} are the same, so these functions are proportional.
Normalisation conditions (item \ref{point-norm-h} in Lemma  and item 
\ref{point-norm} in Definition)  imply they are equal.
\end{proof}

\begin{Rem} \label{rem:dSG}
Equation \eqref{eq:discrete 1D Toda} is identical with a discrete 
version of sine-Gordon equation considered in \cite{KWZ}. 
It is different from the standard form obtained by Hirota.
As a consequence, cellular automata proposed here are different from those 
investigated in \cite{BBGP}.
\end{Rem}

\subsection{A $\tau$-function for the discrete 1D Toda equation}

Using reduction condition for the wave function $\psi$ 
we can derive $\tau$-function satisfying the discrete 1D Toda equation.

\begin{Prop}  \label{prop:KP-to-SG} 
Let $h$ be the function as in Lemma \ref{lem:KP-to-SG}.
Let  $\delta_1$, $\delta_2$ and $\delta_3$ denote the respective  first 
coefficients of local
expansions of $h$ in parameters $t_1$, $t_2$ and $t_3$ at $A_1$, $A_2$ and $A_3$, 
\begin{equation*}
h=t_1(\delta_1 + \dots), \qquad h=t_2(\delta_2 + \dots), \qquad    h={\frac{1}{t_3}}(\delta_3 + \dots).
\end{equation*}
Then the function
\begin{equation} \label{eq:tau-tau-tylda}
\tilde\tau= \tau \, \delta_1^{-n_1(n_1-1)/2}(-\delta_2)^{-n_2(n_2-1)/2}\delta_3^{n_3(n_3-1)/2}
\end{equation}
satisfies the discrete 1D Toda equation \eqref{eq:discrete 1D Toda}.
\end{Prop}

\begin{proof}
Expansions of equation \eqref{eq:KPtoSG-constr-psi} at points $A_1$, $A_2$ and $A_3$ give
$$ T_1T_2T_3^{-1}\rho_1 = \delta_1 \rho_1, \quad 
T_1T_2T_3^{-1}\rho_2 = - \delta_2 \rho_2, \quad
T_1T_2T_3^{-1}\rho_3 = \delta_3 \rho_3. $$
The functions 
$$ \tilde\rho_1= \delta_1^{-n_1}\rho_1, \quad 
\tilde\rho_2= (-\delta_2)^{-n_2}\rho_2, \quad
\tilde\rho_3= \delta_3^{n_3}\rho_3, $$
satisfy
\begin{equation} \label{constr-rho}
T_1T_2T_3^{-1} \tilde \rho_i = \tilde \rho_i, \quad i=1,2,3. 
\end{equation} 
These functions define new potential $\tilde \tau$  related to $\tau$ by \eqref{eq:tau-tau-tylda}.
The new $\tilde\tau$-function satisfies dKP equation and due to condition \eqref{constr-rho} it fulfils
condition \eqref{eq:KPtoSGconstr}. 
As a consequence $\tilde \tau$ satisfies equation \eqref{eq:discrete 1D Toda}.
\end{proof}

\subsection{Explicit formulas for genus $g=0$.}

In the case of projective line the function $h$ reads
\begin{equation*}
h(t) = \frac{(t-A_1)(t-A_2)}{(t-A_3)}.
\end{equation*}
Then,
\begin{equation*}
\delta_1 =\frac{(A_1-A_2)}{(A_1-A_3)}, \qquad \delta_2 =\frac{(A_2-A_1)}{(A_2-A_3)},
 \qquad \delta_3={(A_3-A_1)(A_3-A_2)}.
\end{equation*}

\noindent The formula \eqref{eq:tau-tau-tylda} and the determinant formula \eqref{eq:tau-N-soliton} 
allow to find pure $N$-soliton solutions of the discrete 1D Toda equation over finite
fields.

\section{Examples} \label{sec:Ex}

\begin{Ex}{The vacuum solution of the discrete 1D Toda in  $\FF_5$} 

First, in Figure \ref{fig:SG_vac}, we present $0$-soliton $\tilde\tau$-function 
taking values in the finite field $\FF_5$. 
\noindent 
We fix parameters of the solution from $\FF_5$ as follows: $A_1=0$,  $A_2=2$,  $A_3=3$. 
For this example the Remark at the end of section \ref{rem:periods}
and formula \eqref{eq:tau-tau-tylda}  imply  relations
$T_1^4 \tilde\tau = (\delta_1)^{-2} \tilde\tau$ 
and
$ T_3^4 \tilde\tau = (\delta_3)^2 \tilde\tau $.
Because $\delta_1 = 4$ and $\delta_2=3$ 
the period in $n_1$ variable is equal to 4, and in $n_3$ variable is 8. 
It is also evident from the picture that the reduction constraint
$T_2 \tilde \tau = \gamma T_1^{-1}T_3 \tilde\tau $ where $\gamma =2$ is satisfied. 
Notice also there are no values $0$ in a vacuum solution.

\medskip

\noindent Elements of $\FF_5$ are represented by:

\begin{center}
{\leavevmode\epsfysize=0.35cm\epsffile{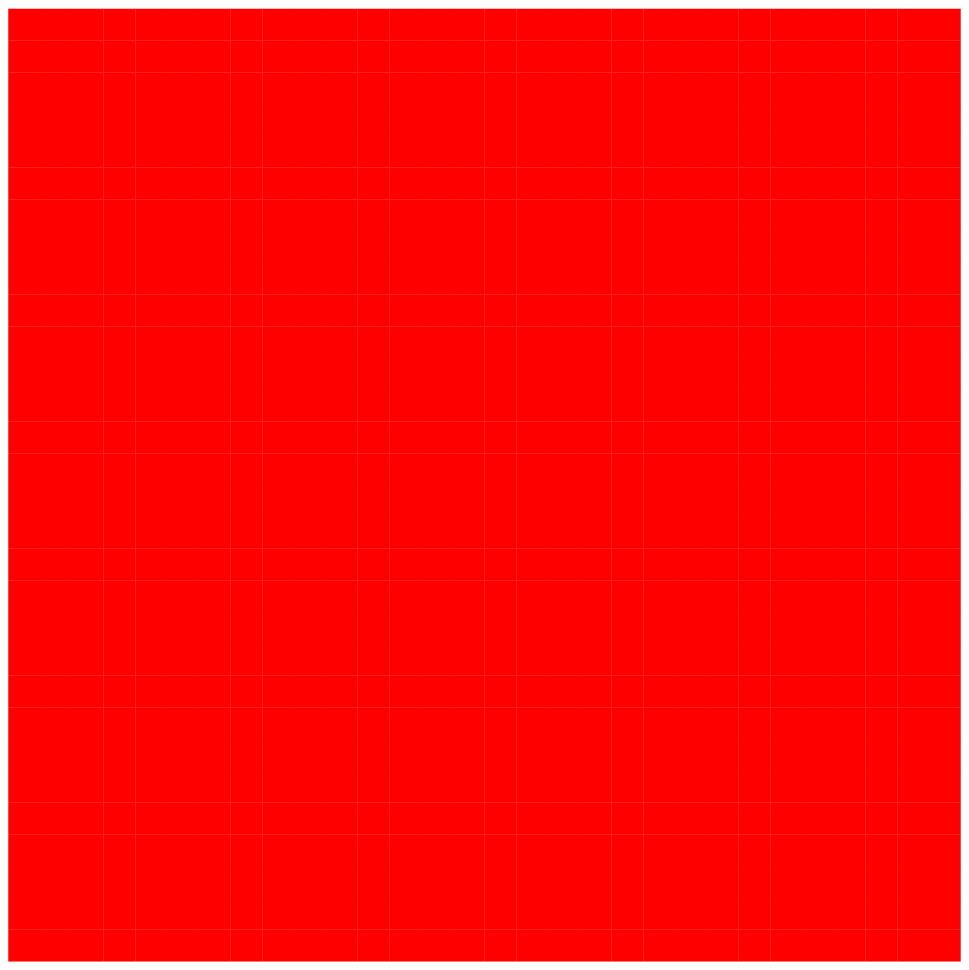} -- $0$, \ 
 \leavevmode\epsfysize=0.35cm\epsffile{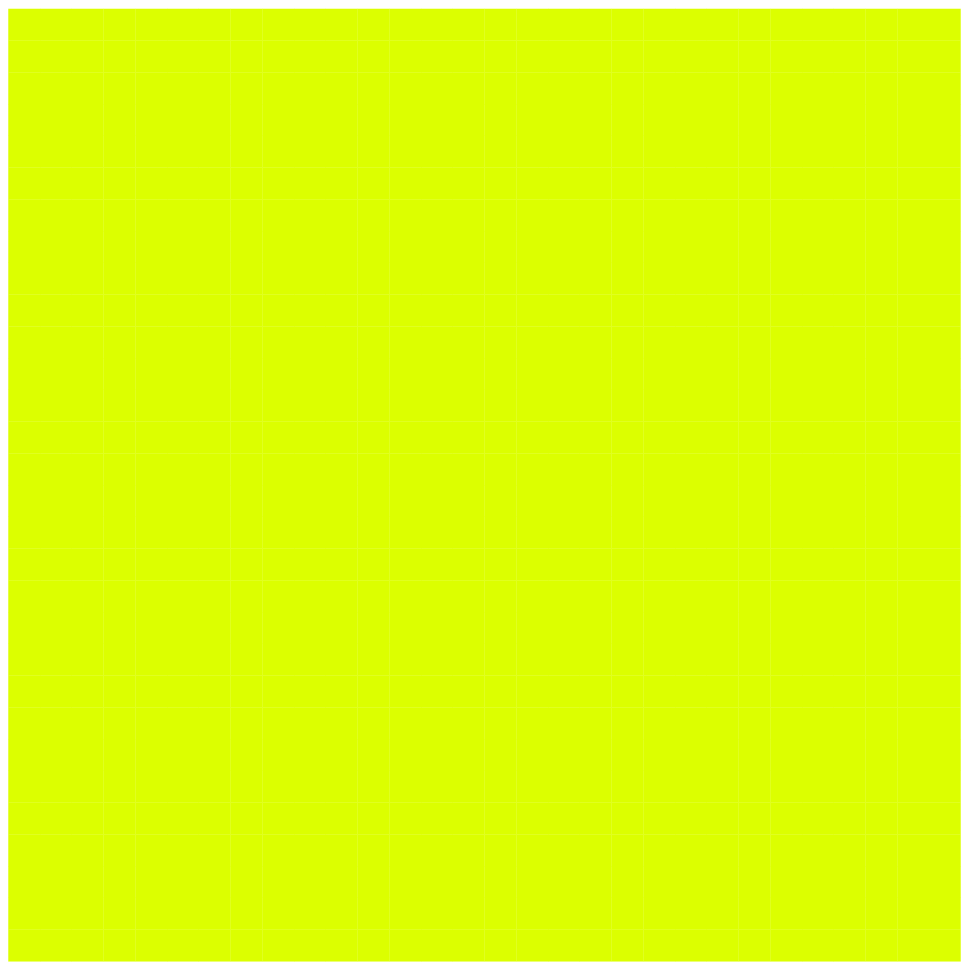} -- $1$, \
 \leavevmode\epsfysize=0.35cm\epsffile{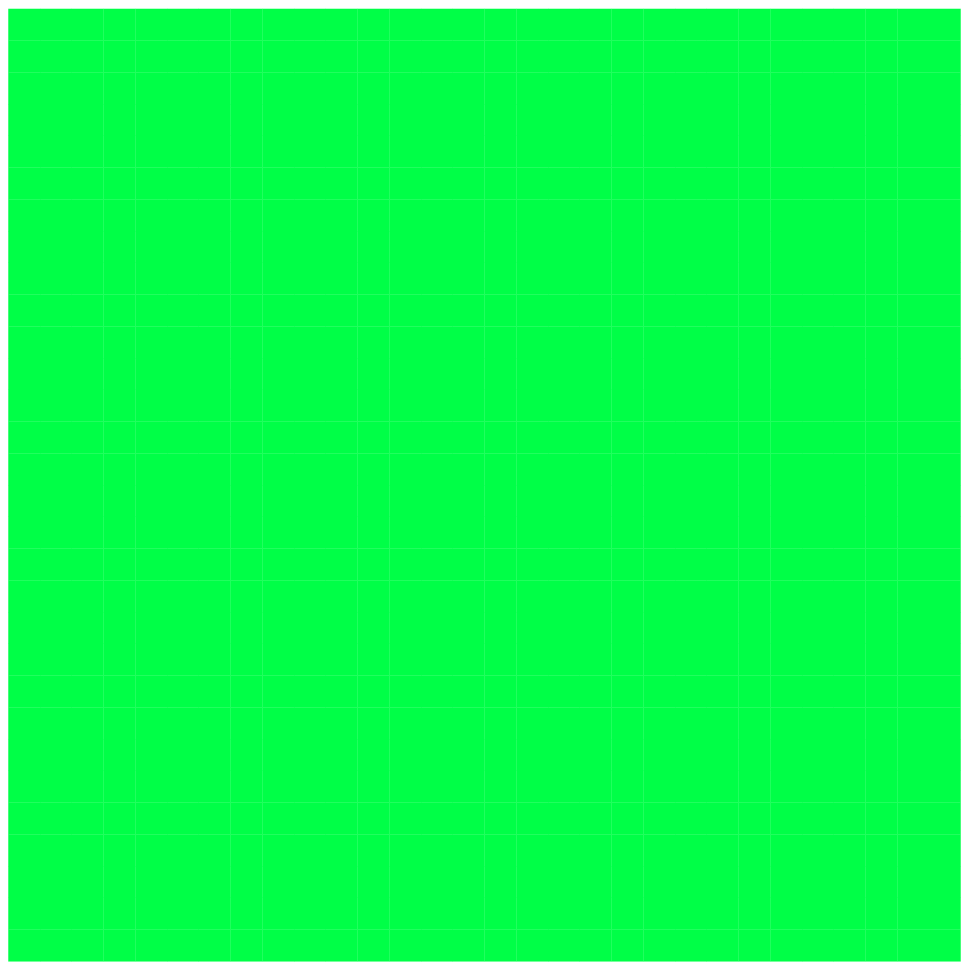} -- $2$, \ 
 \leavevmode\epsfysize=0.35cm\epsffile{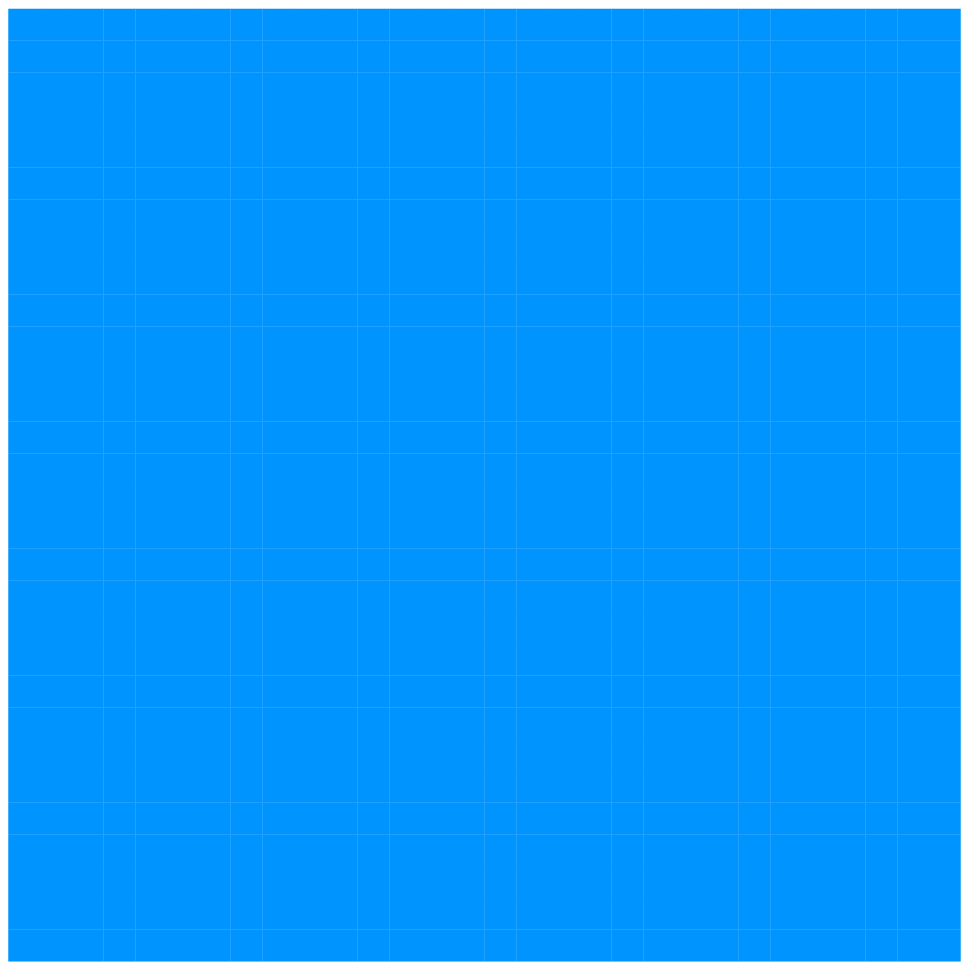} -- $3$, \
 \leavevmode\epsfysize=0.35cm\epsffile{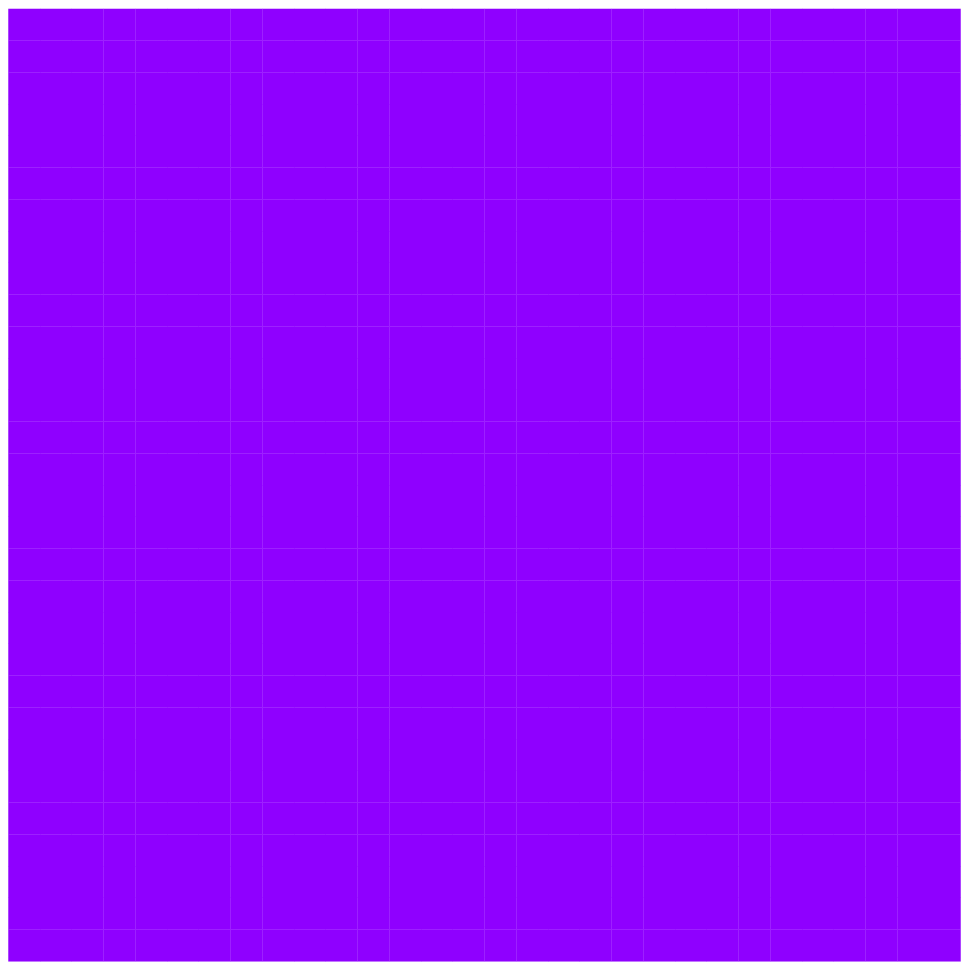} -- $4$.}
\end{center}

\vspace{1mm}

\begin{figure}[!ht]
\begin{center}
\leavevmode\epsfysize=6.1cm\epsffile{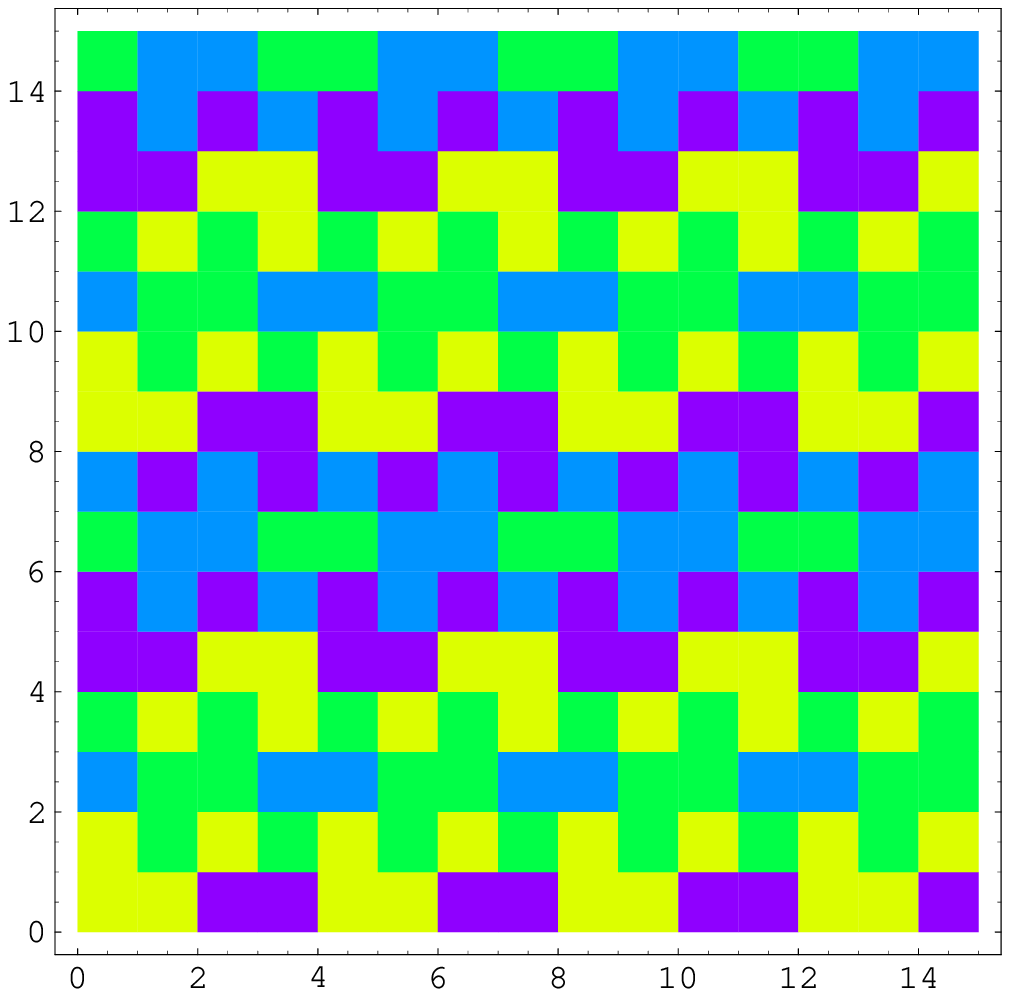} \hspace{0.2cm}
\leavevmode\epsfysize=6.1cm\epsffile{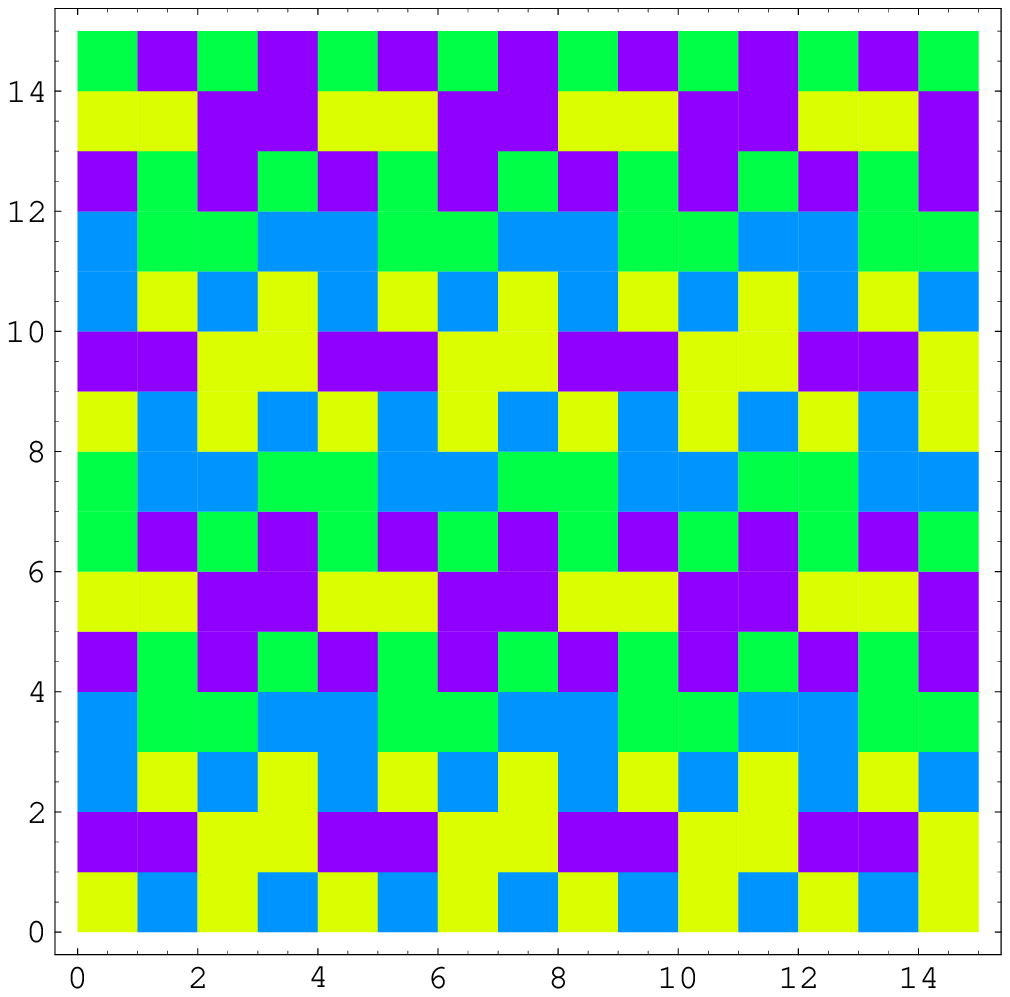}
\end{center}
\caption{
The vacuum solution $\tilde \tau(n_1,n_2,n_3)$ of the discrete 1D Toda equation for $n_2=0$ and $1$. 
Axes: $n_1$  directed to the right (from $0$ to $14$),
 $n_3$ directed upward (from $0$ to $14$).
}
\label{fig:SG_vac}
\end{figure}

\end{Ex}

\begin{Ex}{The breather solution of the discrete 1D Toda in  $\FF_5$}  

The next example, presented in the Figure \ref{fig:SGbreather}, 
is an analog of a breather solution taking values in the finite field $\FF_5$.
Parameters of the solution are chosen in a "symmetric" way from $\FF_{5^2}=\FF_{5}[x]/(x^2+x+1)$ i.e. 
a quadratic algebraic extension of the field $\FF_5$ by the polynomial $(x^2+x+1)$. 
An element $ax+b \in \FF_{5^2}$ is denoted by $(ab)$. 
The adequate Frobenius automorphism is denoted by $\sigma_F$.
Parameters are fixed as follows:\\
$A_1=(00)$,  $A_2=(02)$,  $A_3=(03)$; $C_1=(10)$,  $C_2=\sigma_F(C_1)=(44)$; \\
$D_1=(12)$, $E_1=\sigma_F(D_1)=(41)$,  $D_2=(24)$, $E_2=\sigma_F(D_2)=(32)$; \\
\noindent The $\tilde\tau$-function is normalised by $\tilde\tau(n_1=1,n_2=0,n_3=1)=1$.
Respective periods in variables $n_1$ and $n_3$ are equal to $12$ and $24$.   
The reduction constraint $T_2 \tilde \tau = \gamma T_1^{-1}T_3 \tilde\tau $ 
is  satisfied (for $\gamma =2$) as visible in the figure.

\noindent Elements of $\FF_5$ are represented like in the previous example by:

\begin{center}
{\leavevmode\epsfysize=0.35cm\epsffile{f5-0.eps} -- $(00)$, \ 
\leavevmode\epsfysize=0.35cm\epsffile{f5-1.eps} -- $(01)$, \
\leavevmode\epsfysize=0.35cm\epsffile{f5-2.eps} -- $(02)$, \ 
\leavevmode\epsfysize=0.35cm\epsffile{f5-3.eps} -- $(03)$, \
\leavevmode\epsfysize=0.35cm\epsffile{f5-4.eps} -- $(04)$.}
\end{center}

\vspace{1mm}

\begin{figure}[!ht]
\begin{center}
\leavevmode\epsfysize=6.1cm\epsffile{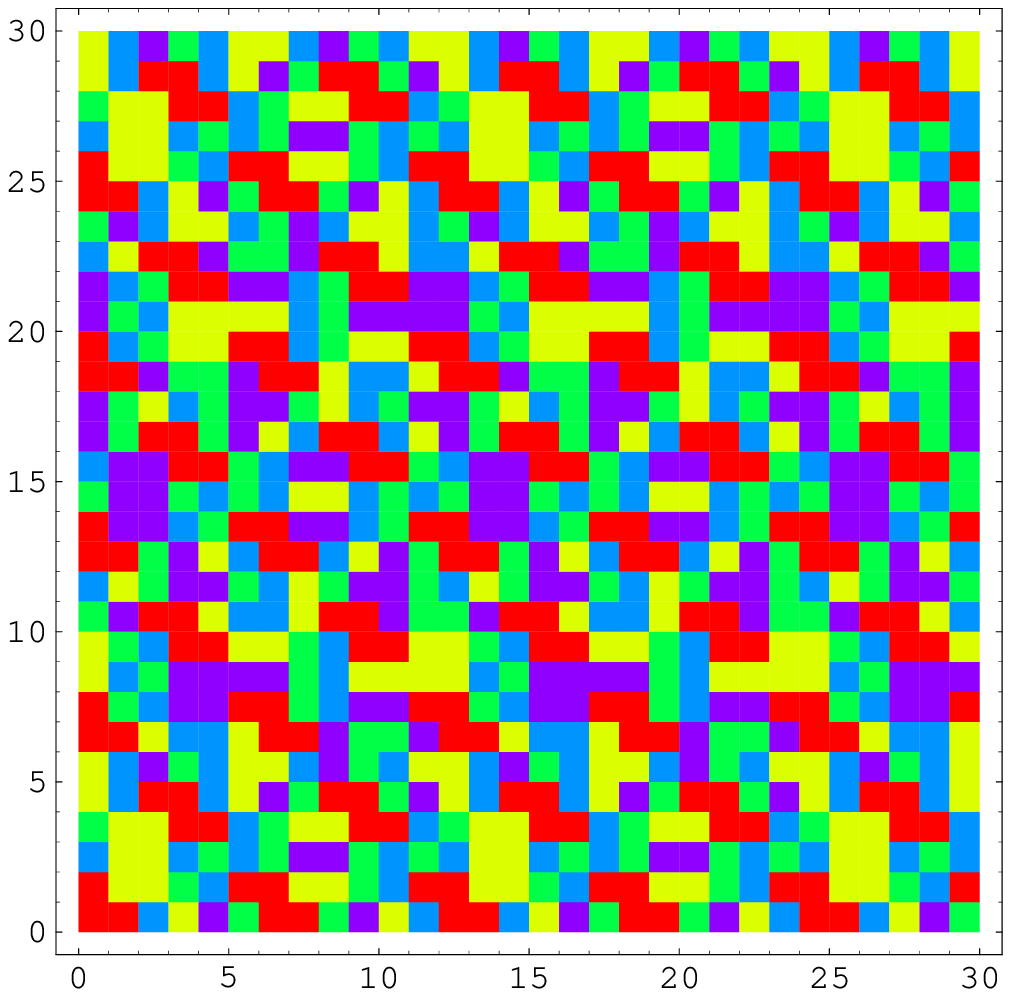} \hspace{0.2cm}
\leavevmode\epsfysize=6.1cm\epsffile{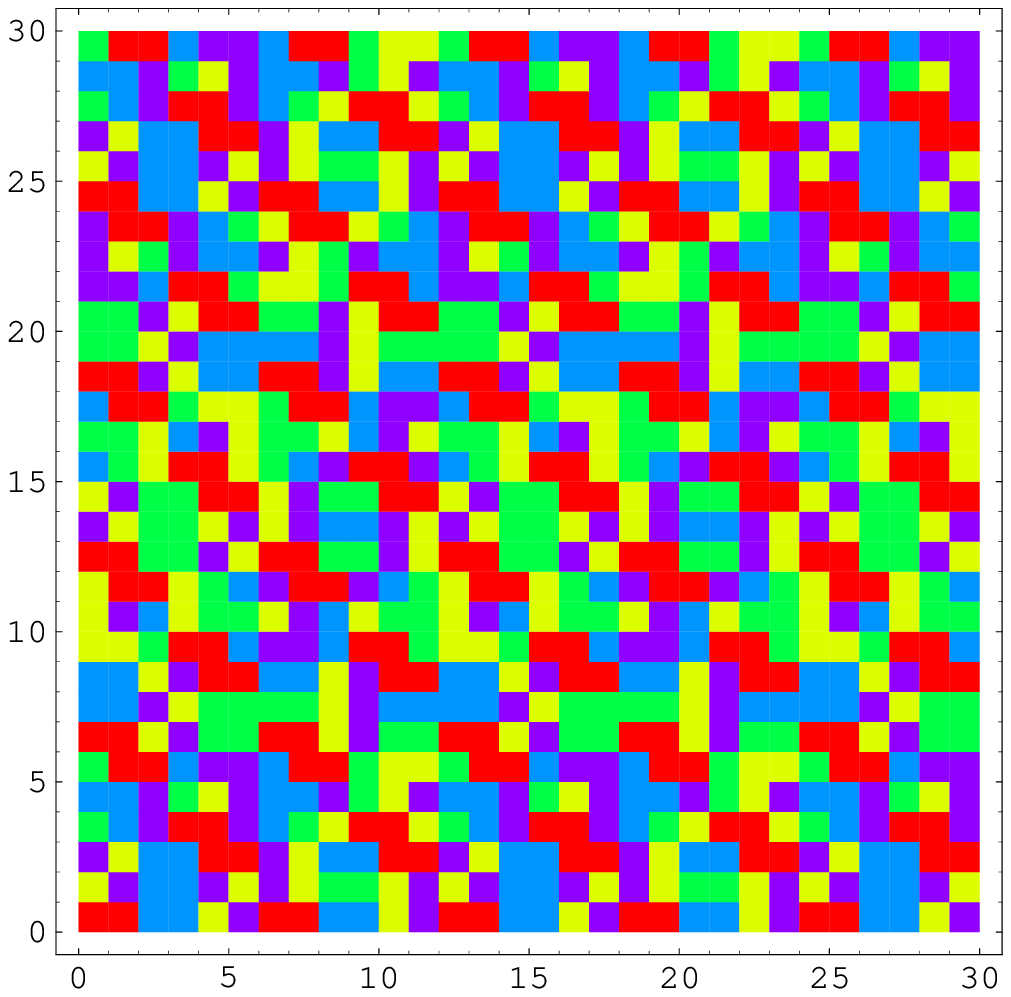} 
\end{center}
\caption{
The breather solution $\tilde \tau(n_1,n_2,n_3)$ of the discrete 1D Toda equation for $n_2=0$ and $1$.
Axes: $n_1$  directed to the right (from $0$ to $29$),
 $n_3$ directed upward (from $0$ to $29$).
}
\label{fig:SGbreather}
\end{figure}
\end{Ex}

\section*{Acknowledgements}
The author would like to express his thanks to prof. Adam Doliwa for discussions and cooperation.
This work is partially supported by Polish Ministry of Science and Information Society Technologies 
(Grant no. 1~P03B~01728).


\end{document}